\documentclass[acmlarge]{acmart}

\makeatletter
\newcommand{\myconfshort}{\acmConference@shortname}
\newcommand{\myconffull}{\acmConference@name}
\newcommand{\myconfdate}{\acmConference@date}
\newcommand{\myconfloc}{\acmConference@venue}
\AtBeginDocument{
  \fancypagestyle{firstpagestyle}{
    \fancyhead{}%
    \fancyfoot[C]{}%
  }
  \fancyhf{}
  \fancyhead[LO]{\@headfootfont\shorttitle}%
  \fancyhead[RE]{\@headfootfont\@shortauthors}%
  \fancyhead[LE]{\@headfootfont\footnotesize \myconfshort, \myconfdate, \myconfloc}%
  \fancyhead[RO]{\@headfootfont\footnotesize \myconfshort, \myconfdate, \myconfloc}%
  \fancyfoot[C]{}%
}
\makeatother
\acmBooktitle{\conffull\@ (\confshort), \confdate, \confloc}

\usepackage{tabularx}
\usepackage{array}
\usepackage{multirow}
\usepackage{booktabs}
\usepackage{caption}
\usepackage[normalem]{ulem}
\useunder{\uline}{\ul}{}
\newcommand{\bumpup}{\vspace*{-.5cm}}

\newcolumntype{L}{>{\raggedright\arraybackslash}X}  
\newcolumntype{W}{>{\raggedright\arraybackslash}p{0.18\linewidth}}

\copyrightyear{2026}
\acmYear{2026}
\setcopyright{cc}
\setcctype{by}
\acmConference[FAccT '26]{The 2026 ACM Conference on Fairness, Accountability, and Transparency}{June 25--28, 2026}{Montreal, QC, Canada}
\acmBooktitle{The 2026 ACM Conference on Fairness, Accountability, and Transparency (FAccT '26), June 25--28, 2026, Montreal, QC, Canada}
\acmDOI{10.1145/3805689.3812408}
\acmISBN{979-8-4007-2596-8/2026/06}

\begin{document}

\title{Co-Designing Organizational Justice Indicators for Algorithmic Systems}


\author{Fujiko Robledo Yamamoto}
\email{furo0108@colorado.edu}
\orcid{0000-0002-7964-0830}
\affiliation{%
  \institution{University of Colorado, Boulder}
  \city{Boulder}
  \state{CO}
  \country{USA}
}

\author{Nicholas Mattei}
\email{nsmattei@tulane.edu}
\orcid{0000-0002-3569-4335}
\affiliation{%
  \institution{Tulane University}
  \city{New Orleans}
  \state{LA}
  \country{USA}
}

\author{Pradeep Ragothaman}
\email{pradeepr@kiva.org}
\orcid{0009-0000-5866-6794}
\affiliation{%
  \institution{Kiva}
  \country{USA}
}

\author{Robin Burke}
\email{robin.burke@colorado.edu}
\orcid{0000-0001-5766-6434}
\affiliation{%
  \institution{University of Colorado, Boulder}
  \city{Boulder}
  \state{CO}
  \country{USA}
}

\author{Amy Voida}
\email{amy.voida@colorado.edu}
\orcid{0000-0002-2490-2063}
\affiliation{%
  \institution{University of Colorado, Boulder}
  \city{Boulder}
  \state{CO}
  \country{USA}
}

\renewcommand{\shortauthors}{Robledo, Mattei, Ragothaman, Burke, Voida}

\begin{abstract}
Fairness in machine learning is often conceptualized narrowly in comparative, distributional terms. In studying stakeholders' concepts of fairness, we find that this framing is insufficient to capture the full range of issues raised. As an alternative, we propose organizational justice as a framework that subsumes distributional fairness as well as other normative concerns. We conduct a case study of organizational justice relative to personalized recommendation in the context of Kiva Microfunds, a nonprofit micro-lending organization whose mission is to increase financial access for underserved communities across the world. We report on the results of co-design workshops conducted with Kiva employees who are involved in different departments and whose roles often lead them to prioritize normative concerns that are most supportive of the stakeholders with whom they work most closely. We apply organizational justice to understand design trade-offs among different normative goals stakeholders invoke. Based on these goals, we identify a suite of metrics that Kiva employees can use to monitor and assess the recommender system's impact on their organizational justice concerns and to seed discussions within the organization about appropriate configuration and deployment of this system in context.
\end{abstract}



\received{20 February 2007}
\received[revised]{12 March 2009}
\received[accepted]{5 June 2009}

\maketitle

\section{Introduction}

There is little argument that algorithmic systems should be fair and that fielded systems often fall short of that standard. However, as Mulligan et al. \cite{mulligan2019thisthing} note, fairness is an "essentially contested concept," one whose definition cannot be narrowly defined. It is unsurprising, then, that multiple interpretations of this concept may be at play in sociotechnical systems \cite{abdollahpouri2020multistakeholder, dodge2019explaining, saxena2019fairness, smith2023scoping}, and that different interpretations of fairness can be in direct conflict \cite{colquitt2015measuring, lee2018understanding, smith2023scoping}, or can interact in complex ways \cite{rahwan2018society, smith2023many}. Apart from a few notable case studies, including the well-studied COMPAS case noted in \citet{mulligan2019thisthing}, we know very little about how these multiple interpretations of fairness interact in real world sociotechnical contexts. In what ways do different interpretations of fairness interact? What are the tradeoffs among different interpretations of fairness? And how might an organization go about understanding the impact of multiple interpretations of fairness as systems are used over time? 

Most research involving fairness in algorithmic decision-making systems has focused on assessing the distribution of outcomes to ascertain whether or not something is fair \cite{lee2019procedural, juijn2023perceived, fu2025implementing}. While fair outcomes are important, they are also insufficient. It is also important to consider and capture other normative concerns that are related to what is fair or just, such as whether a process is consistently applied (procedural justice) and whether the stakeholders involved are treated appropriately (interactional justice) \cite{colquitt2013organizational, colquitt2001dimensionality, greenberg1987procedural}. This more expansive understanding of fairness is referred to elsewhere as organizational justice: a multi-dimensional construct that describes how fairness is applied in workplace decision-making \cite{adamovic2023organizational}. The construct was originally developed in the study of employer—employee relationships \cite{greenberg1987procedural}, but can also be used as a lens to understand other contexts in which an individual is subject to the decisions of a system, including computing contexts (see also \cite{lee2019procedural, juijn2023perceived, fu2025implementing}). 

We are particularly interested in algorithmic fairness in the context of recommender systems: personalized applications that support information access and decision making. Recommender systems serve a variety of stakeholders both inside and outside of the organizations that deploy them \cite{abdollahpouri2020multistakeholder}. Research on fairness in recommender systems has highlighted the complexity of defining and operationalizing fairness in these systems \cite{ekstrand2022fairness} at the conceptual level: what makes a set of recommendations fair? How are tradeoffs among different stakeholders considered in making such an assessment? These are thorny research questions and many scholars choose to ignore them and accept one of a handful of widely-used fairness measures and evaluation frameworks in determining that their algorithmic innovation is fair or at least fairer than what came before.

We approach recommender system fairness from a contextually-situated perspective, asking what a more expansive understanding of fairness—organizational justice—might mean for system design in a particular context. In particular, we explore how this more expansive set of stakeholder concerns can be reflected in metrics and indicators that illuminate how well (or poorly) organizational justice goals are being met. We ask the following questions:

\begin{itemize}
\item RQ1: How do recommender systems stakeholders operationalize organizational justice concerns in ways that extend beyond conventional, distributive fairness?

\item RQ2: How might we operationalize these forms of justice in ways that could 
be monitored in an operational setting?

\item  RQ3: How might we identify a key set of metrics or indicators of justice that will be most conducive to fostering design dialogue among stakeholders over time?
\end{itemize}

We conduct this research in the context of an online recommender system design project with a nonprofit community partner whose mission centers around equity and justice. Kiva Microfunds is a US-based microlending organization with staff around the globe; their mission is to “expand financial access to help underserved communities thrive” \cite{kivawebsite}. Kiva provides an ideal case for understanding the myriad ways in which stakeholders understand what is fair and just. While Kiva employees are all passionate about ensuring that their work is fair and just, they work closely with a diversity of stakeholders around the globe, leading them to prioritize different operationalizations of fairness tailored to those with whom they work most closely. Previously, researchers have characterized the diverse ways that Kiva employees understand fairness in their work context  \cite{smith2023many}. In this research, we use organizational justice theory to help make sense of that breadth of fairness concerns. 


\section{Research Context}
Kiva Microfunds is a U.S. nonprofit organization (501(c)(3)) that uses microlending to provide access to capital, especially in the global contexts in which individuals are financially underserved. As of early 2026, Kiva had lent \$2.4 billion to 6 million borrowers in 79 countries, with 90\% of borrowers reporting an increase in the quality of their life. The 2.1 million Kiva lenders who have supported these loans find them and select them via Kiva’s online microlending platform. Overall, Kiva lenders have seen a 96.4\% repayment rate~\cite{kivawebsite}.

The Kiva ecosystem consists of multiple stakeholders: (1) \textbf{lenders} who are individuals who lend money via the Kiva platform, (2) \textbf{lending partners} who are non-governmental organizations or microfinance institutions in the borrowers’ local communities who support the borrowers throughout the loan application process, (3) \textbf{borrowers} who are the individuals and groups who receive the loans funded at Kiva, and (5) \textbf{Kiva} itself, which has as its mission the alleviation of financial inequity around the world while keeping money flowing through their online marketplace. As an organization, Kiva must make decisions about where and how recommendations are made, how borrower stories are shared, and how lending partners are showcased. 

\begin{figure}
    \centering
    \includegraphics[width=0.66\linewidth]{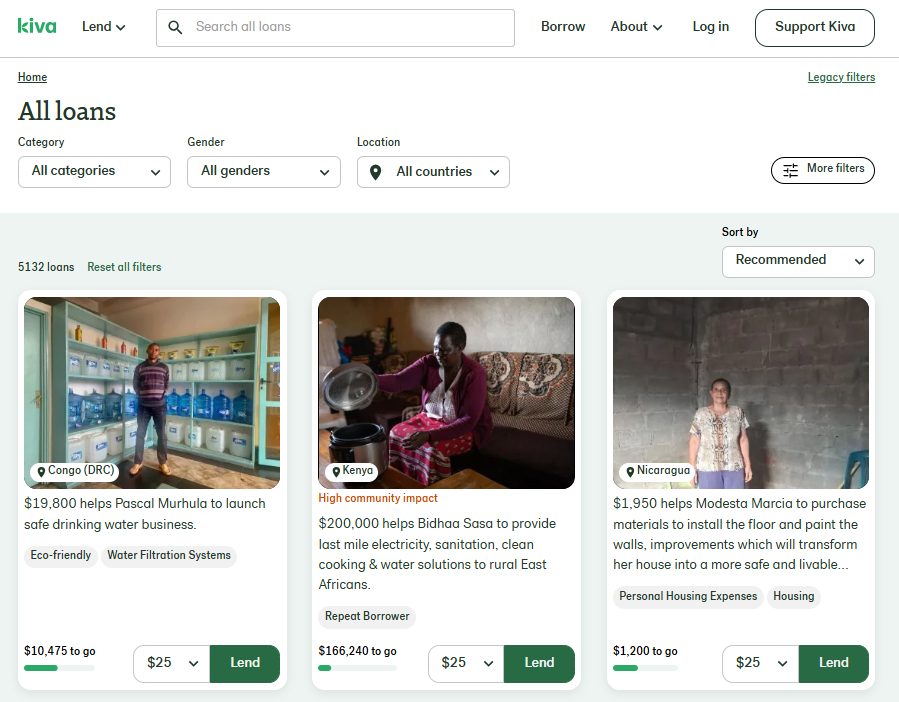}
    \caption{Kiva search options when visiting the loan page.}
    \label{fig:kivaSearch}
    \bumpup
\end{figure}

When a lender visits the Kiva website for the first time, they can use various filters  to search for a lending opportunity that they want to support (Figure \ref{fig:kivaSearch}). While the specific filters change over time and with current events, some of the most stable filters include the borrower’s country or gender identity, the kind of work being supported (e.g., agricultural, educational, artistic, etc…), as well as characteristics of the loan status that are popular with many lenders (e.g., loans that are ending soon, loans that are almost funded, or loans that are matched by corporate sponsors). The lender can specify how much they want to contribute (\$25 is the minimum loan and the most common). Once the loan has been repaid, Kiva notifies the lender and encourages them to relend those funds to a new loan opportunity. 

Kiva provides loan recommendations to lenders in a variety of contexts, including personalized search results based on features of loans they have funded previously, in recommended loan carousels, as well as personalized recommendations in targeted emails to lenders who have enough repaid funds in their account to relend. These recommendations are intended to keep money cycling through the Kiva ecosystem (and increasing Kiva’s overall impact) by appealing to lenders’ known preferences. However,  Kiva also encourages lenders to diversify the kinds of loans they are funding via featured loans and impact ratings that foreground the likely economic or community benefit of a loan.  

Crucially, Kiva has been and continues to iterate on both the algorithmic inputs and the user experience of recommendation largely because the context offers such a challenging design space—both because social justice contexts have historically been some of the most fraught contexts for recommendation systems and because of the diversity of interpretations of fairness across stakeholders. 

Our research team has been collaborating with Kiva for approximately the past five years to help design fair recommendation systems for their platform. Because equity is core to Kiva’s mission, this nonprofit organization is an ideal partner for community-based research exploring the design of fair recommender systems. Kiva is committed to promoting financial equity and improving the quality of life of their borrowers and the communities in which they live, a motivation which requires an understanding of how fairness is being experienced and operationalized in their recommender systems. Employees throughout the organization consider fairness, equity, and justice regularly in their work. As such, Kiva  serves as an ideal critical case for understanding fairness “in the wild.” Our Kiva collaborators have helped co-design the research and provided entrée to employees across the organization for recruitment. We have shared all of our findings with them so they can incorporate the aspects of the research that are most actionable into their internal systems. The findings from this research will help inform the development of a dashboard that displays key justice indicators to foster dialogue related to fairness. 

\section{Related Work}
\subsection{Fairness in Recommender Systems}

As algorithmic decision-making becomes more pervasive, researchers have stressed the need to develop fair and equitable systems to ensure that societal inequities are not replicated by these systems \cite{ekstrand2022fairness, abdollahpouri2020multistakeholder, chaney2018algorithmic, dodge2019explaining}. Recommender systems play a central role in decision-making by shaping which content users see, which introduces the likelihood of design practices that could lead to disparate treatment across the platform \cite{dodge2019explaining}. In platforms like Kiva, recommender systems influence which loans are shown to which lenders, thereby shaping borrower visibility, lending behavior, and funding outcomes. Because of the multisided nature of recommender systems, unfairness can arise in many ways, making it necessary to analyze and examine fairness from various perspectives \cite{burke2017multisided}. 

There have been recommendations for how to integrate fairness into the design of recommender systems \cite{lee2017human, abdollahpouri2020multistakeholder, burke2018balanced, lee2019procedural}. This need has been emphasized by research that represents recommender systems as multistakeholder environments, with impacts across a range of stakeholder groups \cite{abdollahpouri2020multistakeholder,lee2017human, ekstrand2022fairness, burke2018balanced, burke2025decentering}. However, these studies have also highlighted the difficulties with defining and representing fairness \cite{abdollahpouri2020multistakeholder, dodge2019explaining, saxena2019fairness, smith2023scoping}, especially since fairness is highly context-specific. There may be some situations where a degree of unfairness towards some stakeholders is acceptable, especially when it increases the relative fairness for the rest of the group. Each system designer must determine which trade-off is appropriate for their specific use case, a process that is challenging in itself \cite{rahwan2018society}. Furthermore, while implementers may rely on mathematical operationalizations of fairness, these may not correspond to the stakeholder experience of fairness \cite{colquitt2015measuring, lee2018understanding}. \citet{lee2017algorithmic} have argued for integrating “the human element” into definitions of fairness. Operationalizing this, though, is challenging, as perceptions are influenced by many individual, social, and cultural factors \cite{colquitt2015measuring, lee2018understanding, lee2017algorithmic, robert2020designing}. Within the Kiva context, researchers have identified myriad interpretations of fairness across Kiva stakeholders \cite{smith2023many, burke2022multi, sonboli2021fairness}.


Co-design and other participatory methods are a promising way for understanding and evaluating the multi-dimensional effects of fairness in recommender systems \cite{ekstrand2025recommending,smith2024recommend}. Researchers have noted the importance of working directly with stakeholders who influence the system in some way, whether they be the direct users or auxiliary users that affect or are affected by the system in some way \cite{amershi2014power, lee2019webuildai, rahwan2018society, smith2024recommend, sonboli2021fairness}. Stakeholders can often challenge assumptions of traditional machine learning and better inform the development of these systems \cite{amershi2014power, lee2019webuildai, lee2017human}. For example, Smith et al. \cite{smith2024recommend} conducted focus groups with providers to develop fairness metrics for social media content creators and users of a dating app. In this case study, the researchers were able to use the providers' lived experience to ground the operationalizations of fairness. Fairness is context-dependent, therefore requiring that we determine how fairness trade-offs and metrics manifest in specific settings. 

\subsection{Organizational Justice Theory}
Due to the multi-faceted nature of fairness, we draw on \textbf{Organizational Justice Theory}, which provides a way to analyze a breadth of normative fairness concerns—beyond distributional fairness—taking into consideration three dimensions of justice \cite{colquitt2013organizational, colquitt2001dimensionality, greenberg1987procedural}: \textbf{interactional justice}, whether people were treated and represented fairly; \textbf{procedural justice}, whether the decision-making process to arrive at the decision was fair; and \textbf{distributive justice}, whether the outcome was fair. Justice therefore is not just about outcomes, but also about how they are perceived by those involved and about how those outcomes influence one another \cite{colquitt2015measuring, lee2018understanding, colquitt2013organizational}. 

 Human-centered scholars have incorporated dimensions of organizational justice theory into the design of fair and equitable machine learning systems in a variety of contexts, particularly in hiring\cite{lee2019procedural, juijn2023perceived, fu2025implementing, pyle2024us}. For example, Juijn et al. \cite{juijn2023perceived} utilized organizational justice theory to identify trade-off between fairness criteria in algorithmic hiring. They extend fairness not only to include outcomes but also emphasize the importance of providing clarity surrounding how decisions are made. Pyle et al. \cite{pyle2024us} used an organizational justice analytical frame to identify distributive, procedural, and interactional injustices when using emotion AI in hiring interviews, particularly for marginalized groups. They strongly recommended against adopting the technology in hiring due to its potential to exacerbate inequities and increase emotional labels for marginalized folks.  

Organizational justice theory has also been applied in other contexts to evaluate the fairness of machine learning systems. Lee et al. \cite{lee2019procedural} presented a procedural justice framework for decisions involving resource allocations among people and groups. They emphasized the importance of transparency and control when designing algorithmic decision-making systems. Fu and Lyu \cite{fu2025implementing} used organizational justice to evaluate the fairness and ethical perceptions of using facial recognition technology. The framework allowed them to identify and balance the needs and perspectives of multiple stakeholders and to call-out the consequences of how a lack of transparency was violating users’ expectations of fairness.  These examples highlight the benefits of organizational justice theory for analyzing recommender systems since it helps foreground multiple stakeholder perspectives, helps weigh trade-offs between costs and benefits, and helps reflect the dynamic nature of fairness. Similarly, we use this framework to weigh the design trade-offs suggested by the participants—taking into account the multiple variants of fairness and related justice indicators that can serve as collaborative tools and that focus on factors beyond outcomes. 

\section{Method}
\subsection{Participants}
We conducted remote, co-design workshops with 9 Kiva employees from 5 different departments and who each had specific stakeholder expertise (i.e., depending on their role, each participant had specific knowledge about fairness issues that may impact the borrower, lending partner, or lender). We conducted 3 workshops; each workshop consisted of two, one-hour sessions. Each workshop had 2 to 4 participants from individuals from different departments to ensure diversity of thought. We conducted these workshops between December 2024 and June 2025. We received IRB approval to conduct this research and participants were compensated for their participation.

We centered Kiva employees as the primary stakeholders for this research as they will be the stakeholders making collective decisions about next steps for the justice interventions that Kiva makes in its algorithms and platform. Because of their different roles at Kiva, employees typically understand fairness in different ways that correspond to the external stakeholders with whom they work most closely. Because they work at Kiva, they also understand the tradeoffs that are so fundamental to Kiva’s mission of financial equity. As such, this broad base of Kiva employees are ideal for understanding what metrics and indicators will be most valuable to Kiva in assessing the impact of various justice interventions and in negotiating next steps.

The participants had worked at or with Kiva an average of 5.5 years, ranging from 1.5 years to 10 years. Some participants worked directly with system design and engineering and had a more precise technical understanding of the current state and capabilities of existing systems. These participants were, however, much more at arms length from key stakeholders and so their participation in co-design workshops was complemented by employees who were more involved in front-end user experience and interface design (typically representing the interests of lenders) and employees who interfaced more with people than technology (e.g., recruiting and supporting lending partners around the world or working with corporate donors.)  Co-design workshops were tailored to foster dialogue across expertise and perspectives.

We recruited participants across multiple Kiva teams. Some of these participants had already participated in a different study and had expressed interest in participating in the workshops. We also worked with our research partner at Kiva who shared our recruiting email. We utilized a pre-workshop interest form to determine who was interested in participating, to schedule sessions, and to collect information on demographics and areas of expertise. The participants were compensated \$20 for their participation in the workshops. 

Because of participant scheduling constraints (workshops were conducted during the Kiva workday), we were not always able to schedule workshops with the range of expertise that we would have preferred. Nevertheless, all workshops facilitated conversation among multiple stakeholders and our analysis spanned all workshops, enabling us to contrast different perspectives that might not have been present in the same workshop.


\subsection{Co-Design Workshops}
The goal of the workshops included: (1) identifying which factors are relevant for fairness and identifying design trade-offs and affordances of prioritizing certain factors and the implications for each stakeholder. The workshop content and structure was designed with input from the partnering organizations to ensure that their needs were represented. Based on the preferences of Kiva leadership, we split the workshop into two one-hour sessions: A and B.  We identified three goals for each session, which were organized into phases of stage setting, design activities, and discussions of how Kiva might assess algorithmic fairness. 

We conducted the sessions on Zoom and shared all participants into a shared set of Google Slides, which were designed by the research team and contained all the activities of the workshop. Each participant had edit access and was able to complete the activities directly on the Google Slides. When participants preferred to discuss rather than write directly on the Google Slides, the facilitator took notes and verified with participants that important points were accurately captured. 

\subsubsection{Stage Setting} 
We organized stage setting activities in both workshops to ease the participants into the topic of fairness. For Workshop A, we asked the participants to introduce themselves and to share a story about a time when they were worried about something being fair in their work with Kiva, including what happened, who was involved, and what was the ultimate outcome. This helped set the stage for talking about fairness and also helped generate an initial list of factors that influence fairness. For Workshop B, we set the stage by giving participants scaffolding and asking them to brainstorm any factors that they think are important when considering fairness that were not generated in Workshop A. This list was referenced and updated throughout Workshop B.

\subsubsection{Design Activities} 
The design activities in each workshop centered around two fairness scenarios: search results and personalized emails. The Workshop A design activity focused on how to achieve fairness in the context of Kiva’s curated search results for a new Kiva lender. This scenario required participants to consider the tradeoffs in fairness factors in the absence of data about lenders’ prior search history. The Workshop B design activity focused on how to achieve fairness in the context of Kiva’s curated recommendations for existing Kiva lenders via personalized emails. This second scenario opened the discussion to an additional suite of fairness factors related to the lender. 

For each design activity, we presented participants with a hypothetical but realistic scenario in which they had to maximize fairness for either the borrower or lender.  We asked the participants to state all of the factors that they would take into consideration when maximizing fairness in each scenario. We then asked them to pick the three most important fairness factors and discuss how they would weigh each relative to the others and how prioritizing each factor would be likely to impact each stakeholder. Finally, we asked the participants to describe how either the search results or personalized emails would look given their prioritization of the three factors, including what information they would contain. 

\subsubsection{Fairness Assessments}
For the final part of each workshop, we asked questions to explore how fairness might be assessed: 
How do you know when you have achieved fairness for [stakeholder]? Which fairness consideration do you think would be the biggest “win” for lenders but the highest “liability” for borrowers?
Which fairness consideration would be the biggest “win” for borrowers but the biggest “liability” for lenders? The participants discussed the metrics that they felt would indicate the highest level of fairness, the midpoint level of fairness, and the failure to achieve fairness for a specific stakeholder. The participants also reflected on the workshop process. 

\subsection{Data Collection and Analysis}
We audio recorded and then transcribed the workshops. In total, we collected 360 minutes of audio (60 minutes for each workshop). We also collected the Google slides on which each workshop collaboratively completed their activities (126 slides in total). Our data also includes field notes taken by the workshop facilitator. To analyze the data, we conducted inductive and iterative coding, drawing on grounded theory techniques for open coding \cite{strauss1990basics}. 

For the first phase of analysis, the first author inductively open coded the transcripts and created memos exploring possible themes and any other observations. Initial open codes included different loan attributes that impact fairness, considerations/tensions of fairness, barriers to fairness, and design suggestions. These initial codes and memos were shared and discussed with the rest of the research team. 

For the second phase of analysis, the first author identified preliminary themes inductively from the data. The preliminary themes included:  (1) lenders come to Kiva with assumptions and with hopes of making emotional connections, (2) there is a tension between what is most impactful versus what is most engaging, (3) there are cultural and contextual misunderstandings that impact fairness perceptions, and (4) fairness is dependent on which stakeholder is prioritized. These inductive codes are predominantly clustered around three larger themes: relational elements (i.e., interactional justice), the importance of providing context and explanations (i.e., procedural justice), and differential experiences with outcomes (i.e., distributive justice).  The research team then engaged with existing theoretical frameworks to interpret and refine these themes, finding that Organizational Justice Theory provided a meaningful lens for understanding the patterns in the data. This abductive approach, which included moving iteratively between empirical findings and theoretical frameworks, led us to re-examine the data through the three dimensions of Organizational Justice Theory. Rather than imposing these categories a priori, we used them to sharpen and organize themes that had already emerged from the data. 



\section{Findings}
In this section, we discuss the design considerations and trade-offs for each type of justice. 

\subsection{Interactional Justice}
\textbf{Interactional justice} refers to the relational aspects of justice \cite{colquitt2013organizational, colquitt2001dimensionality}. For Kiva participants, interactional justice manifests in the way that borrowers are portrayed in the descriptions of loan opportunities and in how Kiva lenders are presented with information and choices about loan opportunities. The participants introduced two design trade-offs: (1) eliciting emotional engagement versus introducing too much bias and (2) providing helpful context versus providing too much information. 

\subsubsection{Design Consideration: Representing Borrowers on the Platform}

The participants consistently emphasized the importance of \textit{“centering human dignity”} (P4) in Kiva’s platform, which ties directly into how information and stories are shared. This commitment is not only a moral value, but it is also seen as essential for lender engagement and retention.  However, curating content on Kiva involves balancing emotional engagement with potentially increasing the bias that influences lending behavior. As one participant explained: \textit{I think whenever I fund a loan, there's usually a rational component which is like, is this a good business idea?... And then an emotional component [...] do I feel something?}  (P9)

As P5 noted, photographs are especially important: \textit{“photos are the humanizing elements that enable the person looking at the story to understand a bit about what that person’s experience is [...] and to identify with some part of their experience”} (P5). This work of humanization is inherently curated and could increase the amount of bias in the process. Even the aesthetics of a picture can affect how a lender perceives the borrower. For example, P8 states how bias can be introduced through pictures:\textit{If you show people 3 images, and one of them has a really clear, large picture of someone's smiling face, I would guess they'd be more likely to choose that one [...] that's not really fair, because we're priming them.} (P8)


\subsubsection{Design Consideration: Providing the Right Amount of Information}
A key component of interactional justice is clear communication, which signals respect and care by providing enough context to make sense of the differences between the types of loans that are shown, including the loan attributes. However, providing all of this information to truly understand the borrower’s situation and the loan attributes introduces another tension: balancing context without overwhelming lenders. Lenders need to have knowledge regarding Kiva’s model, the borrower’s cultural and socioeconomic context, and the potential short and long term impact of the loan on the borrower’s life.

The participants emphasized the tension between providing enough context…

\begin{quote}
 \textit{We've found lenders might want to look at like a couple of data points when they make a decision. And they're not able to take in an entire economic, country level analysis…it's a tension that we've tried to address, but it's not something that I would say we've resolved.} (P5)
\end{quote}

… without providing so much information that lenders become overwhelmed: \textit{We ended up having to make a collapsible borrower page, so that the first thing you see is a more bullet-pointed version of all that information, and you can totally uncollapse it and read it all. But a lot of people apparently don't....} (P6)

Furthermore, the language used becomes an important part of humanizing stories and invoking empathy. Creating mutual understanding, however, is challenging: \textit{How do we streamline even the way we define things….For example, everybody thinks climate in a certain way and the first ideas that come up is issues like agriculture, solar…We need to incorporate that in in our website...} (P9) 


To support more informed and empathetic decisions, several participants suggested highlighting the additional benefits that accompany some Kiva loans. For example, \textit{“financial education, business training”} (P2) or \textit{“access to a bank account, access to some sort of [health] insurance”} (P1). P2 added that for certain sectors, such as agriculture, there may be additional services offered along with the loan, such as: \textit{...agricultural training, provide access to the right seeds, the best seeds, training on climate resilience…We would want to like highlight, that they're not just getting a loan; they're actually getting a lot of other support.”} (P2)

In summary, these two tensions—emotional engagement versus even more bias and meaningful context versus information overload—highlight the delicate design balance required to uphold interactional justice on the Kiva platform. Respecting borrowers’ dignity and supporting lenders’ understanding are deeply intertwined goals, which may pull in different directions. The choices Kiva makes in how it represents borrowers, communicates loan details, and structures the user experience are not merely aesthetic or operational; they are ethical decisions that shape how respect, empathy, and agency are experienced on the platform. As such, maintaining interactional justice requires ongoing reflection on whose voices are heard, how stories are told, and what assumptions are challenged or reinforced in the process. 

\subsection{Procedural Justice}
\textbf{Procedural justice} refers to the consistency, transparency, perceived legitimacy, and autonomy in the processes and decision-making procedures used to arrive at outcomes. For the Kiva participants, autonomy, in particular,  emerged as a recurring theme when discussing fairness. Participants questioned the role of Kiva in influencing loan outcomes, weighing the tensions between Kiva ordering loan opportunities by one or more loan attributes (e.g., time left to be funded or amount left to be funded) versus allowing lenders to filter and sort as they preferred. The participants also weighed the tension between promoting lenders’ autonomy by providing a diversity of possible loans to fund versus personalizing a focused subset of loans that a lender sees based on their prior interactions on the site. 

\subsubsection{Design Consideration: \textit{“Playing God”} versus \textit{“Playing the Market”}}
The participants raised questions about the degree to which Kiva should make decisions about what metadata should be incorporated into decisions about curating and ordering loans, for example, in search results. These procedural questions had a significant impact on how participants perceived system fairness. Unsurprisingly, participants also acknowledged a variety of different ways that Kiva could use metadata and that each of these uses impacts procedural justice. For example, P7 suggested that sorting loans based on metadata about the time remaining in the fundraising period could be a way to intervene to help ensure that all loans are given visibility at the end:

\begin{quote}
\textit{The time remaining on a fundraising period…. So this is one thing, I think, that should factor into fairness, because we understand the sort of funding curve of loans. }
\end{quote}


Another participant recommended using metadata about loan visibility similar to a farmer’s market, with a deliberate rotation of loans to ensure that they all have visibility on the site: 

\begin{quote}
\textit{They'll have vendors move to different stalls so people get the best places at different times of the year, in order to try to increase fairness for all of the vendors in a market…because everybody gets the same amount of days, no matter how much money that loan is for.} (P1) 
\end{quote}


However, the participants cautioned against being too prescriptive  in curating loan recommendations for lenders. P4 likened this to \textit{“playing God,”} stating that even though ensuring equitable distribution is important, lenders should also be able to have a choice in determining what is fair and important to them: \textit{How much do we want to play God in terms of like, telling people what impact is and what fair is versus say, providing them with all the necessary filters and tools on their page, to let them find the ones that they think are the most impactful and sort of most deserving. It's a very challenging one to figure out.}

Balancing identifying and promoting certain loans as opposed to letting the market decide what gets funded is challenging: how does the organization decide which borrowers need more help in promoting their loans, and in turn, is this help fair for other borrowers? What processes does the organization have to decide who to promote and are these clear and consistent? 

\subsubsection{Design Consideration: Broad versus Focused Selection of Loans}

The participants also highlighted design tradeoffs related to lenders' autonomy in the context of personalized recommendations. Autonomy, or the ability to have some control over the decisions made by a system, is an integral factor of promoting procedural justice \cite{lee2019procedural, juijn2023perceived}. When loan recommendations are personalized based on prior lending behaviors, P8 worries that the resulting \textit{“echo chamber”} would undermine lender autonomy, limiting their lending choices (i.e., displaying only loans from certain sectors, countries, etc.). P3 explained it as follows: \textit{“There's so many things now where everything is so customized, and I enter this tunnel, like, I don't ever get a chance to see something that maybe doesn't lie outside my typical lending parameters. And sometimes that wild card is like, ‘Yeah, I'll go for that.’”} P8 posed the question that gets at the crux of this design tradeoff:

\begin{quote}
\textit{How do you help a lender go on an educational journey and explore different kinds of loans? I think a lot about, like the type of person that comes to Kiva. I actually think they are curious. They do care about these things, and so, if we can…help them see other opportunities to make an impact in a way where they're able to learn and then make that choice as opposed to it being, like, forced on them. To me, that would be a fairness-enhancing change.}  (P8)
\end{quote}

However, the participants also discussed how showing loans that are most similar to prior loans, might also help retain lenders, which in turn, could result in more loans being funded. \textit{Converting the most lenders, or like getting people to, you know, support loans in the end, like really is in the interest of the borrowers, because the more loans that we fund, the that means more borrowers get funding…but you can't really do that without privileging some loans over others} (P5).

Since the more loans funded increases the number of borrowers who are given an opportunity for increased economic autonomy, this tradeoff results in a clear tension between which stakeholders’ autonomy to prioritize. To address this tension, P4 called for a more transparent process by which lenders are given autonomy—not just through a wider breadth of lending opportunities—but in deciding what type of lending experience they want to have. He suggests the following:
\textit{I almost feel like you need to choose your own journey when you come to the site of like:  I want Kiva to choose for me or I want to choose myself. And then that dictates the path you go down, like here all the filters, so that you can refine your search versus Kiva will serve up options for you.} (P4)

In summary, these tensions around personalization and transparency reflect the complex procedural choices that shape user experience and fairness on Kiva. Procedural justice, in this context, is less about who gets funded and more about \textit{how decisions are made}, how much autonomy is prioritized for which stakeholders, and how prioritizing certain variables and/or experiences impacts fairness perceptions. 

\subsection{Distributive Justice}

\textbf{Distributive justice} refers to the resulting allocation of resources across individuals or groups, and whether those outcomes are perceived as appropriate given needs, contributions, or context. In order to evaluate whether an outcome is fair, there needs to be some expectation of what the outcome should be, even if that expectation varies among stakeholders.  The relative fairness of these expected outcomes also needs to be assessed against the broader context of other possible outcomes. For Kiva participants, this translated to questions about how lending outcomes are distributed across different stakeholder demographics, countries, and risk profiles to determine who is benefiting and who is being overlooked. 


\subsubsection{Design Consideration: Balancing Quantified Outcomes with Perceptions}

The main tension in distributed justice is how to best evaluate the outcomes, whether tracking quantified data (e.g., rates of funding per country) or by whether there is a \textit{perceived sense} of fairness, especially since each loan is different and each stakeholder has their own perceptions and expectations of the experience: \textit{If we treat every loan the same when every loan is not actually the same, maybe that's not fair to borrowers} (P1).
This leads to a \textit{“big range of outcomes”} (P5) on Kiva that extend beyond whether a loan is funded or not to whether a lender continues to loan to Kiva or whether a lending partner grows their portfolio on Kiva. 

When asked about what kind of indicators would help them evaluate fairness, most of the participants stated that it is difficult due to the varying definitions of fairness. P6 expressed this uncertainty: \textit{“would it even be a measure of fairness to say, oh, well, all of these borrowers  put their loan up on the same day; did one of them fund faster than the other? Or did they all fund at the same rate? Like I mean, is that a measure of fairness?  I'm not sure.”} The participants suggested that there may be certain indicators that could highlight potential fairness issues, such as, the speed of loans being funded (P9), but resolving any fairness issues that such an analysis might identify was understood as an entirely different challenge.


For each workshop, we asked participants to identify three key indicators of fairness, with lender conversion and funding rates of loans that are expiring soon or nearly funded being the most popular selection in the three sets of workshops. While all groups were able to converge around three individual indicators of fairness, rendering an overall assessment of fairness across all indicators proved to be more challenging, as fairness factors are interrelated and no one could predict how moving the needle on one fairness indicator would affect the others: \textit{I guess I would approach it as an experiment, and start, you know, start with weights for each variable  and determine some measures of success, and how that's going, and then I don't know, tinker in favor of one, and see how that changes that kind of thing} (P4).

The participants also emphasized that different fairness indicators would be relevant for different stakeholders: \textit{“there's different impacts to different stakeholders based on the outcome”} (P5). For example, P6 talks about how funding a loan is only part of the equation; if that loan does not get paid back, even though the outcome for the borrower was positive, the lender and lending partner may not feel like the interaction was fair. Some other participants believed that a way to assess fairness to borrowers was really to focus on the outcomes of lenders and lending partners. For example, assessing conversion, or the number of lenders that become recurrent users, is ultimately a positive outcome for the borrower: \textit{Hopefully I can convince [the lender or lending partner] to keep coming back...depositing more and more money, which will ultimately mean that more borrowers would be served in the future} (P8).

Understanding the impact of the outcomes, therefore, requires more than just relying on quantified data. There is a need to also evaluate how the outcome was perceived by the stakeholders. As P3 described, \textit{“it is through the outcome that you strengthen the relationship [with Kiva lenders and lending partners].”} Evaluating perceptions of fairness can be quite tricky. For example, P8 discussed how fairness depends on perception: \textit{Some people had some losses and didn't understand that that was a risk [in lending on Kiva]. And is that a fairness issue? I'm not sure, but, like probably from their perspective it was} (P8).

The participants struggled to articulate how one might measure or collect data about stakeholder perceptions of fairness. P4 summarized it this way: \textit{“There's kind of like an amalgamation of like interpretations of fairness that all kind of come together in different places. Like, taking into account what's fair to lending partners or lenders. We probably need a perspective, like a framework, for thinking about fairness.”}

In summary, distributive justice involves evaluating outcomes to determine if a result is fair. For Kiva, distributive justice requires having both quantified data about which categories of loans are being funded and the ability to assess how stakeholders perceive the fairness of these outcomes. 

\section{Discussion: Organizational Justice Indicators as Boundary Objects}
Organizational justice theory offers a useful lens for identifying and distinguishing among the many design considerations inherent in working towards fairness in a sociotechnical system, teasing fairness apart into three dimensions: interpersonal treatment (interactional justice), processes (procedural justice), and outcomes (distributive justice). Beyond identifying the many different ways that fairness can be understood within a system (e.g.,\cite{smith2023many, smith2023scoping, smith2024recommend, lee2019procedural, juijn2023perceived}), using organizational justice as an analytic lens provides a broader basis for identifying key organizational justice indicators or metrics. Due to the varying interpretations of fairness and the diversity of forms of organizational justice,  this suite of indicators and metrics can serve a critical role as boundary objects among disparate Kiva stakeholders. Boundary objects, which are “objects which are both plastic enough to adapt to local needs and constraints of the several parties employing them, yet robust enough to maintain a common identity across sites” (\cite{star1989institutional}, p. 393), are particularly helpful for addressing challenging problems such as making fair decisions in sociotechnical systems \cite{leigh2010not, star1989institutional, bowker2000sorting, sharma2022cocreating}.  A suite of organizational justice indicators, those that all stakeholders understand—albeit from their own vantage point—can help support dialogue about how current fairness interventions might be improved, where the goal is not complete agreement, but rather, fostering understanding and collaboration in ideating a productive path forward.

In the following sections, we discuss those indicators of organizational justice that best encapsulate the design considerations prioritized by participants. We discuss how surfacing these indicators can support a shared framing of fairness challenges and help Kiva employees begin to converge on next steps for fairness interventions (Table~\ref{tab:boundary}).

\begin{table}[tbh]
\centering
\small
\begin{tabularx}{\textwidth}{%
    >{\raggedright\arraybackslash}X
    >{\raggedright\arraybackslash}X
    >{\raggedright\arraybackslash}X
    >{\raggedright\arraybackslash}X}
\toprule
\textbf{Organizational Justice Element} &
\textbf{Design Consideration} &
\textbf{Proposed Indicator} &
\textbf{Acts as a Boundary Object By…} \\
\midrule

\textbf{Interactional justice: does the platform foster understanding and connection respectfully and clearly? } &
-Eliciting emotional engagement versus introducing too much bias \newline
-Providing helpful context versus too much information &
-Borrower characteristics, such as age, gender, etc. \newline
-Photo characteristics, such as emotion conveyed \newline
-Depth of information provided to lenders \newline
-Quality of information provided &
Revealing underlying assumptions guiding interactions on Kiva \\

\midrule

\textbf{Procedural justice: are algorithms transparent, consistent, and justifiable?} &
-“Playing God” versus “playing the market” \newline
-Providing a broad versus focused selection of loan opportunities &
-Loan characteristics, such as time remaining, category, amount requested, etc. \newline
-Percent of users who search versus those who rely on recommendations \newline
-User perceptions of algorithmic tools &
Creating clear guidelines as to why and how decisions are being made, thus increasing transparency \\

\midrule

\textbf{Distributive justice: are outcomes equitable across stakeholder groups? How do we know?} &
-Quantified versus perceptual outcome data &
-Risk to lender versus impact to borrower \newline
-Feedback on fairness perceptions &
Mapping outcomes to understand impacts on stakeholders and align on action points \\

\bottomrule
\end{tabularx}
\caption{How fairness metrics can act as a boundary object across the dimensions of organizational justice}
\label{tab:boundary}
\bumpup{}
\end{table}

\subsection{Indicators of Interactional Justice}
Interactional justice emphasizes the relational aspects of fairness. To make interactional justice more concrete, the participants surfaced fairness indicators such as borrower characteristics, photo attributes, and communication styles. These factors can be operationalized into metrics that enable Kiva employees to identify trade-offs related to relational interactions. For example, tracking which borrower features disproportionately affect funding outcomes can reveal whether certain groups are systematically advantaged or disadvantaged in how they are perceived. Identifying concrete fairness indicators that could give rise to bias, such as borrower characteristics or photo attributes, can provide a common reference point for visualizing and discussing fairness trade-offs. For example, Kiva employees who work closely with borrowers could bring up concerns related to the difficulties of being able to take a quality photo for their profile, while Kiva employees who work closely with lenders could talk about how lenders tend to respond better to certain types of images. The multiple viewpoints can be surfaced and decisions can be made by taking into consideration all of these factors that impact fairness. The participants also suggested using fairness indicators to determine best communication practices. Other researchers have indicated how challenging it can be to find the balance of how much and what type of  information to convey to ensure that people are empowered to make choices when interacting with recommender systems \cite{rader2018explanations, dodge2019explaining, rader2018explanations}. In the context of Kiva, too little information can reduce borrowers to statistics, taking away the much needed personal touch to help form connections between borrowers and lenders and can lead to confusion as to why certain borrowers were recommended to them. Too much information, however, can be overwhelming and unhelpful. By tracking the level and depth of information provided and its impact on engagement, Kiva employees can obtain a better understanding of communication practices that promote fairness, clarity, and dignity for all stakeholders. 

\subsection{Indicators of Procedural Justice}
Procedural justice emphasizes that decisions should be consistent, transparent, and give stakeholders a sense of control \cite{abdollahpouri2020multistakeholder, lee2017human}. The participants raised a key question: should Kiva act as a neutral facilitator of lending (to the extent that is possible) or as an active agent shaping fairness outcomes? Would these decisions lead to more equal distribution of loans that are funded? Do algorithmic suggestions increase the breadth of loans that users are exposed to? How are stakeholders perceptions of fairness influenced by increasing autonomy or oversight on algorithmic recommendations? This tension plays out in how the participants imagine the platform could make recommendations. To decide the role that Kiva should play, the participants identified several indicators: loan characteristics, tracking users who rely on algorithmic recommendations, and user perceptions of algorithmic tools. Taken collectively, these indicators could help provide a sense of whether Kiva should promote or prioritize certain loans that seem to be less popular, or whether to provide a broad or a focused selection of loan opportunities. 

For example, some participants suggested that Kiva could randomize the order of loans or allow users to adjust how much autonomy they have when selecting loans versus relying on algorithmic suggestions. In practice, we could envision justice metrics, such as the percentage of users engaging with algorithmic suggestions versus those who have a more directed interaction, to act as a boundary object in dialogue and negotiations around organizational influence versus user autonomy. Kiva employees could also use indicators, such as loan characteristics (i.e., time left to fund, category), to discuss which types of loans need additional support on the site and therefore should be recommended to new and returning lenders. Envisioning these metrics as boundary objects could help not only make decisions, but also communicate to other stakeholders how and why these decisions were made. 

\subsection{Indicators of Distributive Justice}
The most straightforward way to evaluate distributive justice is by examining how outcomes are allocated across categories of people, contexts, or resources \cite{colquitt2013organizational, colquitt2001dimensionality, greenberg1987procedural}. This process involves identifying patterns in comparative data to determine if something is fair or not. For example, one unfunded loan tells us little, but, a number of consistently unfunded loans from certain borrower groups, regions, or sectors signal a level of systemic inequity. In addition, the question of fairness in distribution depends not only on the outcome itself, but also on what outcomes matter most to each stakeholder (i.e., repayment rates, risk level, categories of loans funded, benefits of lending partners). These outcomes could clash in certain situations, revealing the need to view these metrics as boundary objects that help articulate goals and tensions to move close towards alignment among stakeholders.

In this case, we could envision using indicators, such as funding rates based on borrower demographics, loan types, region, or lending partners, and subjective perceptions of fairness to map outcomes for stakeholders. Creating a map of outcomes can be helpful to identify possible points of leverage \cite{sharma2022cocreating}, that is, areas where Kiva employees can intervene or influence. These indicators could also help coordinate action within Kiva by introducing the questions: Who is responsible for creating changes? How will these be implemented? When will they be reassessed? What have we learned from these outcome metrics that can inform future decisions or processes? Outcome metrics then, instead of being static variables, become actionable and informative for how to improve fairness for stakeholders. 

\subsection{Putting it All Together}
Fairness challenges are prevalent in sociotechnical systems; these problems are complex, uncertain, and contexted \cite{wanzenbock2020framework, sharma2022cocreating, palavicino2023co}. In this research, our aim was to identify indicators that could represent the different dimensions of organizational justice within Kiva’s sociotechnical system. These indicators are not designed to be considered individually, but rather, holistically. As indicated by the participants and prior research, fairness is multidimensional and cannot be evaluated from a single vantage point \cite{colquitt2015measuring, lee2018understanding, lee2017algorithmic, robert2020designing}. Having a suite of indicators that can behave as boundary objects, fostering conversation among a diversity of Kiva employees is a starting point towards the development of algorithmic interventions for fairness. 

\section{Limitations and Future Work}
In the ongoing  collaboration with Kiva, future research will involve using these insights to build and deploy an organization justice dashboard for Kiva employees to use to monitor a breadth of justice indicators that serve as boundary objects, spurring dialogue about how various indicators seem to influence each other and how to balance fairness concerns across stakeholder groups. 

We acknowledge the importance of including multiple stakeholders in co-design research. In this work, we relied on the participants to surface and represent concerns related to other stakeholder groups with whom they interact regularly. While involving other stakeholders, such as lenders or borrowers, would provide valuable perspectives, doing so presented substantial challenges. For privacy reasons, the research team does not have access to lender information, making recruitment difficult. Engaging with Kiva’s global and diverse borrower population poses additional logistical, linguistic, and ethical complexities. 

Future work will be needed to explore ways to meaningfully include additional stakeholder groups, potentially through collaborations with local partners or community organizations that can help facilitate inclusive and context-sensitive participation. Alternately, future research could focus on designing remote, asynchronous mechanisms for other stakeholders to contribute additional justice indicators. Because different stakeholders contribute distinct forms of expertise, these participatory inputs could help iterate on the suite of boundary objects used over time. Integrating participatory mechanisms into the system, however, requires thoughtful preparation and intention \cite{colquitt2015measuring, lee2018understanding}. Without careful attention, participatory feedback loops could inadvertently amplify existing inequities by overrepresenting stakeholders who are most vocal or most available to engage. 

\section{Conclusion}
In this research, we have made the following contributions: (1) Used the organizational justice framework to untangle fairness tensions in recommender systems in a micro-lending institution, (2) Provided tangible recommendations for how to operationalize the three justice elements into indicators to evaluate trade-offs and to serve as a dialogue tool, and (3) Demonstrated the need to expand fairness discourse beyond the distributive aspects of the system (i.e., outcomes), which are typically the sole focus of the study of fairness in machine learning systems.

Kiva is not just a lending platform–it is an institution that makes decisions, communicates narratives, and shapes outcomes through structured processes. Kiva’s stakeholder ecosystem has distinct needs, perspectives, and ideas about what fairness entails. Due to this multi-faceted nature of fairness, it is imperative to create frameworks that capture and honor its complexity. As researchers have echoed \cite{abdollahpouri2020multistakeholder, dodge2019explaining, saxena2019fairness, smith2023scoping}: there isn’t one approach that will optimize or promote fairness for all stakeholders in decision-making systems. What may work in one setting may not necessarily transfer to a different setting \cite{amershi2014power, lee2019webuildai, rahwan2018society, smith2024recommend, sonboli2021fairness}. Therefore, fairness requires identifying design trade-offs and evaluating the impact of fairness beyond just focusing on the distribution of outcomes. The Organizational Justice Framework is a useful tool to help identify trade-offs and possible fairness indicators that can help seed discussions about how fairness is represented in sociotechnical systems.

\section{Endmatter}
\subsection{Generative AI Usage Statement}
The authors did not use Generative AI to assist in the writing of this publication.

\begin{acks}
We are grateful to all the participants who dedicated their time and expertise in helping us untangle dimensions of fairness. We are thankful to our research partners for providing feedback. This work would not have been possible without the support of the National Science Foundation (Grant no. 2107577). 
\end{acks}

\bibliographystyle{ACM-Reference-Format}
\bibliography{citations}

\end{document}